# Spin Vortex Resonance in Non-planar Ferromagnetic Dots

Junjia Ding, Pavel Lapa,[†] Shikha Jain,[‡] Trupti Khaire, Sergi Lendinez, Wei Zhang, Matthias B. Jungfleisch, Christian M. Posada, Volodymyr G. Yefremenko,[§] John E. Pearson, Axel Hoffmann, and Valentine Novosad[*]

Materials Science Division, Argonne National Laboratory, Argonne, IL 60439, USA

**Abstract**

In planar structures, the vortex resonance frequency changes little as a function of an in-plane magnetic field as long as the vortex state persists. Altering the topography of the element leads to a vastly different dynamic response that arises due to the local vortex core confinement effect. In this work, we studied the magnetic excitations in non-planar ferromagnetic dots using a broadband microwave spectroscopy technique. Two distinct resonance frequency ranges were observed depending on the position of the vortex core controllable by applying a relatively small magnetic field. The micromagnetic simulations are in qualitative agreement with the. experimental results.

[†] On leave from Texas A&M University, College Station, TX 77843, USA.
[‡] Presently at HGST, A Western Digital Company
[§] Presently at High Energy Physics Division, Argonne National Laboratory.
[*] Correspondence and requests for materials should be addressed to V. N. (email: novosad@anl.gov)

**Introduction**

The investigation of spin dynamics in geometrically confined ferromagnets is an important research topic due to both the technological and the basic science relevance. Some prominent examples include the nanoscale spin textures with non-collinear magnetization arrangement such as skyrmions,[1] domain walls,[2,3] and spin vortices.[4] A vortex-type magnetization is energetically favorable for micron and sub-micron disks with no crystalline anisotropy. The long-range dipolar forces govern the spins dynamics in the vortex-state. Of particular interest is a low-frequency excitation mode associated with the vortex core gyration. Besides the magnetization of saturation, the vortex eigenfrequencies depend on the dot thickness-to-diameter ratio.[5] The excitation spectrum depends only weakly on the value of an in-plane applied field because the external field is compensated by the demagnetizing field of the shifted vortex.[6] Furthermore, it has been reported that the intrinsic pinning due to grain boundaries,[7] surface roughness[8] or exchange bias[9] could also influence the vortex core gyration frequency. While the vast majority of the works on the dynamics of spin vortices deal with the geometrically flat elements, here we have studied the circular elements with intentionally altered topography. This was achieved by using a pre-patterned substrate with the characteristic lateral dimensions significantly smaller that the diameter of the desired magnetic element. In our case, the nonmagnetic disks were first defined via lift-off process, and then covered by another patterned ferromagnetic layer with the same geometric center but a larger diameter. As the result, we obtained a "hat-like" structure of the sample, schematically depicted in Figure 1(a). Then the dynamic properties of these non-planar elements were systematically investigated using a broadband microwave spectroscopy and micromagnetic modeling as a function of their dimensions and amplitude of the external field. As we show below, the introduced disk-shaped step (also referred as a vortex barrier) provides a strong geometric confinement and vortex core pinning effect.

**Results**

**Element geometry.** Two sets of samples were fabricated. The first one is a reference sample, a flat Permalloy (Py, $Ni_{80}Fe_{20}$ alloy) disk with diameter of 1 micron, and thickness of 50 nm. The second one is a non-planar, or an "engineered" dot, which is also circular in shape, with the same dimensions but has altered topography. Shown in Fig. 1(a) is a sketch of the cross-section profile of the engineered dot in which the magnetic material at the center has been lifted 25 nm by preparing a non-magnetic (Titanium) island prior to the deposition of the ferromagnetic material. Figures 1(b) and (c) show representative scanning electron microscope images of the non-planar dot with 200 nm diameter of the step and the reference dot. The step height $T_b$ and the dot thickness $T_{Py}$ have been fixed at 25 nm and 50 nm, respectively in all samples presented in this paper, while the inner diameter $D_b$ varied from 150 to 300 nm.

**Dynamic response.** In order to characterize the dynamic response of the samples, a microwave transmission measurement has been performed in a broad frequency range using a vector network analyzer (VNA). Applying an external magnetic field $H_{app}$ allowed us to probe how the dynamic response of the vortex core changes once it has been displaced from the central area of the disk.

Figure 1(d) compares the absorption spectra of the engineered dot taken at remanence ($H_{app}$ = 0), and in applied field ($H_{app}$ = -180 Oe) after saturating the sample with a −1.5 kOe external field. Interestingly, for a higher value of applied field we detect the resonance frequency of the vortex translation mode at ~0.4 GHz, while for the $H_{app}$ = 0 Oe the frequency increases almost 50%, to ~0.6 GHz. The same microwave measurements performed for a flat $Ni_{80}Fe_{20}$ dot (e.g. the reference sample, without vortex barrier) are shown in Fig. 1(e). The frequency change when the $H_{app}$ is decreased from -180 Oe to 0 Oe is insignificant for the reference dot. A slightly higher frequency for shifted vortex state is in fact expected. It originates from additional dipolar and exchange forces due the vortex spin structure deformation. The striking result is that while the resonance frequencies for *the shifted vortex* states are the same for both samples, their values *in remanence* differ significantly. This suggests that the dynamic response of the engineered dot can be fine-tuned by controlling the relative position of the vortex core with respect to the vortex barrier using a relatively small magnetic field. One could speculate that for the engineered dot subjected to $H_{app}$ = -180 Oe, the vortex core is located outside the circumference of the barrier and thus it encounters the dipolar fields averaged within the outer boundary of the entire dot. As the $|H_{app}|$ value is reduced to zero, the vortex core shifts to the dot center and its gyration is now governed by the dominant contribution from the magnetic charges confined within the barrier boundary. As a result, there appears the ~0.2 GHz frequency difference between these two core positions. Obviously, this effect is not observed for the reference dot (Fig. 1(d)) due to the lack of the barrier.

To further clarify the observed effect we fabricated and systematically investigated the dots with variable barrier diameter. Figures 2 (b - e) are the representative absorption spectra taken at remanence ($H_{app} = 0$ Oe) for the 50 nm thick, 1 micron diameter Py dots but with $D_b$ varied from 150 to 300 nm. The vortex mode of all modified dots is characterized by the domination of one peak. The frequency of this main mode ($V_0$) decreases as the barrier diameter is increased. This trend is similar to the results of previous studies[10] for planar $Ni_{80}Fe_{20}$ disks where the resonance frequency was reported to decrease with the increase in the dot diameter (for a fixed thickness of the element). The micromagnetic modeling yields similar results as it is discussed below. The 2D absorption spectra (Fig. 2(a' – e')) also contain important information about the magnetization evolution process. Let us consider the reference dot ($D_b = 0$ nm, Fig. 2(a')) and the dot with $D_b = 200$ nm (Fig. 2(c')) as two examples. The vortex resonance corresponding to the translational mode appears at $H_{app}$ ~250 Oe in both structures, indicating that the value of the vortex nucleation field is not affected by presence of the vortex barrier. Unlike to the almost field-independent spectra of the reference dot, a discontinuous step-like frequency change is observed in the engineered dot. For instance, for $H_{app}$ = -180 Oe and 0 Oe, the frequency difference is ~200 MHz for the dot with $D_b$ = 200 nm (Fig. 1(d)), while it is only ~15 MHz for the reference dot (Fig. 1(e)). As it was mentioned before, this jump in the frequency is attributed to the changes of the position of the vortex core. In a high magnetic field, when $H_{app}$ is just below the nucleation field, the vortex core is located close to the outer edge. As the magnetic field $H_{app}$ is gradually decreased, the vortex core progressively displaces towards to the center of the dot. The intensity of the resonance line in an engineered dot remains unchanged till $H_{app}$ = -150 Oe. In the field range from -150 Oe to -60 Oe the signal disappears (or is below our experimental sensitivity limit). We speculate that in this fields range the vortex gyration is significantly suppressed due to the pinning effect of the barrier edges. (This vortex core pinning field range varies as a function of the barrier diameter as the vertical line indicated in Fig. 2(b' to e').) The resonance reappears again when $H_{app}$ is decreased below -60 Oe suggesting that the vortex core is now fully inside the barrier circumference. Stronger dynamic dipolar fields of the barrier cause the gyrotropic mode frequency shift to a much higher values. The frequency does not change in small positive fields till the gyration stops when the core reaches the barrier edge again. With further magnetic field increase, the vortex core overcomes the barrier border and the low frequency resonance line re-emerges again. Similarly, to the nucleation fields, the vortex annihilations fields almost do not differ in the modified and reference dots. Thus, while we have seen that altering the dot topography has a profound impact on the low-field vortex core dynamics, its overall magnetostatic properties (e.g. the hysteresis loop) remain unaffected. This is different from the case of dot-on-dot structure where bi-stable magnetic states were reported.[11]

**Micromagnetic modeling:** A systematic micromagnetic study has been performed to further investigate the static and dynamic response of the modified dots. The simulations confirm that in spite of such significant alteration to the topography of the disk, the vortex state remains the ground state for the system. Figures 3 (a, b) show the magnetization distribution in remanence for the dot with $D_b$ = 200 nm and in an in-plane magnetic field $H_{app}$ = -180 Oe.

A significant difference in the eigenfrequencies for these two distinct cases (e.g. the core located outside and inside of the barrier circumference) were also confirmed micromagnetically.. Shown in Fig. 3(c) are the calculated relative energy profiles of the engineered dot (triangle symbols) and reference dot (square symbols) plotted as a function of the displacement of the vortex core. A clear difference between the two sets of results indicates the strong effect of the vortex confinement when the core is at the center of the element. The energy profile vs the core displacement can be approximated as a parabolic function[12] $E(X) = E(0) + 1/2\kappa X^2$, where $\kappa$ is the effective stiffness coefficient, $E(0)$ is the energy at the equilibrium position and $X$ is the vortex core displacement. It should be emphasized that the dot self-induced magnetostatic energy of moving vortex provides the dominant contribution to $E(X)$. Using the simulation data shown in Fig. 3(c), one can find the remanent values of stiffness coefficients ($\kappa$) as $0.94 \times 10^{-20}$ J/nm$^2$ and $1.7 \times 10^{-20}$ J/nm$^2$ for the reference and engineered dots, respectively. Since the frequency is directly proportional to $\kappa$,[13] the vortex core gyrates faster when trapped inside the barrier. There appears no significant difference between the energy profiles for engineered (triangle symbols) and reference (square symbols) dots when the vortex core is located outside the barrier ($H_{app}$ = -180 Oe), Figure 3(d). The asymmetry in the energy profile due the vortex structure deformation in the shifted state can be accounted by adding a cubic term.[6]

As the vortex core is very small, the dipolar forces originating from the dynamic magnetic charges outside the vortex core govern its gyration. These charges can be calculated using a so-called "side-surface charges free" analytical model.[14] Within this model, the magnetization distribution of precessing vortex obeys boundary conditions such that there is no net magnetization component perpendicular to the dot's lateral surface. Altering the topography that leads to formation of a step-like barrier will inevitably impose an additional requirement so the magnetic "charges" on the barrier edges surfaces are minimized as well. Figure 3 (e) shows representative images of temporal changes in the divergence of the magnetization for flat and engineered dots (upper and lower images, respectively). It is clear that the volume magnetostatic charges contributing to the vortex dynamics and defining its eigenfrequency are well-distributed across the dot for the reference sample, while their counterparts in the engineered dot are predominantly located in the central area circumscribed by the barrier edge. The same simulation was also performed for $H_{app}$ = -180 Oe as shown in Fig. 3(f). To our surprise, in this case the spatial distribution of changes in the volume charges is almost identical for both samples resulting in a similar vortex gyration frequencies.

Finally, to further understand the effect of the vortex barrier to the translational mode frequency, systematic simulations have been performed as a function of the barrier size. Shown in Fig. 4(a) is the summary of the experimental (square symbols) and micromagnetic (solid line) frequencies as a function of $D_b$. While there is a noticeable quantitative discrepancy between the experimental and computational results, they are in a good qualitative agreement. Figure 4(b) shows the simulated frequency plotted as a function of barrier thickness $T_b$. The frequency continuously increases with $T_b$, it almost doubles in comparison to the reference dot for $T_b$ = 40 nm. These results demonstrate that the vortex resonance frequency can be effectively controlled by adjusting the barrier geometry. Interestingly, the low field experimental and micromagnetic frequencies for engineered dots shown in Fig. 3(a, b) scale universally when replotted as a function the barrier geometric aspect ratio $T_b/D_b$. This is similar to how the translational mode frequency in flat disks scales as a function of the disk thickness to diameter ratio.[10]

## Summary

A nonmagnetic nanodot inserted under a mesoscale $Ni_{80}Fe_{20}$ dot was shown to provide a geometric confinement effect causing the changes in the vortex translational mode frequency. Two distinct resonance frequency ranges were observed depending on the position of the vortex core (inside or outside of the barrier) controllable by applying a relatively small magnetic field. By comparing the experimental data and micromagnetic simulations if was found that the frequency of the gyrotropic mode increases as the thickness-diameter ratio of the barrier is increased. Further studies of such non-planar ferromagnetic elements will be focused on the details of the pinning mechanism, its possible impact on the energetics of the vortex core reversal process and the high frequency spin dynamics.

## Methods

**Sample fabrication.** The engineered dots were fabricated using a multistep electron-beam (EBL) lithography process. First, the disk arrays with diameter in a range of 150 nm to 300 nm and alignment marks were defined on polymethyl methacrylate (PMMA) resist, accompanied by e-beam evaporation and lift-off process of a 25-nm-thick titanium film. The second step EBL patterning of 1-micron diameter disks followed by deposition of 50-nm-thick $Ni_{80}Fe_{20}$ and lift-off completes the fabrication process. The barriers and the disk are concentric as is confirmed by Scanning Electron Microscopy imaging.

**Spectral measurements.** In order to characterize the dynamic properties of the samples, a coplanar waveguide (CPW) with a 3 µm-wide-signal line was fabricated on top of each dot array using optical lithography followed by a Ti(5 nm)/Au(150 nm) sputter deposition and a lift-off process. Microwave transmission measurements have been performed in the 0.05 ~ 10 GHz frequency range using a broadband microwave vector network analyzer (VNA). The microwave transmission was measured by sweeping the frequency for fixed magnetic field. Applying an in plane magnetic field allowed us to probe how the dynamic response of the vortex core changes once it has been displaced from the central area of the disks. Since the focus of this paper is the fundamental gyrotropic vortex mode, all results are presented in the field range of -500 ~ 500 Oe and frequency range of 0.05 ~ 1.0 GHz . Prior to the magnetic field sweep, the samples were magnetized at the 1.5 kOe field.

**Micromagnetic simulation details.** Systematic micromagnetic modeling was performed using mumax3 code.[15] Typical parameters for $Ni_{80}Fe_{20}$ (the saturation magnetization $M_s$ = 700 kA/m, the exchange constant $A$ = 13×10$^{-12}$ J/m, the damping factor 0.01, the gyromagnetic ratio $\gamma$ = 2.8

GHz/kOe, negligible magnetocrystalline anisotropy) and 5×5×5 nm$^3$ cell size were used in the simulation.

**Acknowledgements**

This work was supported by the U. S. Department of Energy (DOE), Office of Science, Basic Energy Sciences (BES), under Award # DE-AC02-06CH11357.

**Figure Captions**

**Fig. 1**: (a) A sketch of the dot with $D_b$ = 200 nm with the simulated remnant magnetization state of the engineered dot. The x-component of the magnetization is represented from red to blue in a range of +1 to -1. Scanning electron micrographs of the dot with (b) 200 nm barrier diameter ($D_b$) and (c) reference dot. (d) and (e) shows the experimental FMR curves with $H_{app}$ = -180 Oe and 0 Oe for $D_b$ = 200 nm and 0 nm, respectively

**Fig. 2**: (a-e) The experimental remanent FMR absorption curves of the dots with $D_b$ = 0 nm (reference dot), 150 nm, 200 nm, 250 nm and 300 nm. The corresponding 2D FMR absorptions spectrums are shown in (a' to e'). The vertical lines in (b' to e') indicate the field range that vortex core jump is pinned at the barrier.

**Fig. 3**: The simulated magnetization state (central layer) of the dot with $D_b$ = 200 nm for $H_{app}$ = 0 Oe (a) and -180 Oe (b). (c) and (d) shows the plots of the simulated energy profile of both the reference dot and the engineered dot as a function of the core displacement for $H_{app}$ = 0 Oe and -180 Oe. Solid lines are the fitting results of the profile with the parabolic function. The simulated magnetic charge density ($\nabla \cdot M$) changes when the core is displaced from the equilibrium position (when it is performing gyrotropic movement) for the reference dot and the engineered dot with $H_{app}$ = 0 Oe (e) and -180 Oe (f). The color code is in log scale.

**Fig. 4**: (a) and (b) shows the simulated and experimental results of the gyrotropic mode frequency of the modified dots as a function of the barrier diameter ($D_b$) and the barrier thickness ($T_b$), respectively. All results have been reformatted and plotted as a function of the barrier aspect ratio ($T_b/D_b$) as shown in Fig. 4(c).

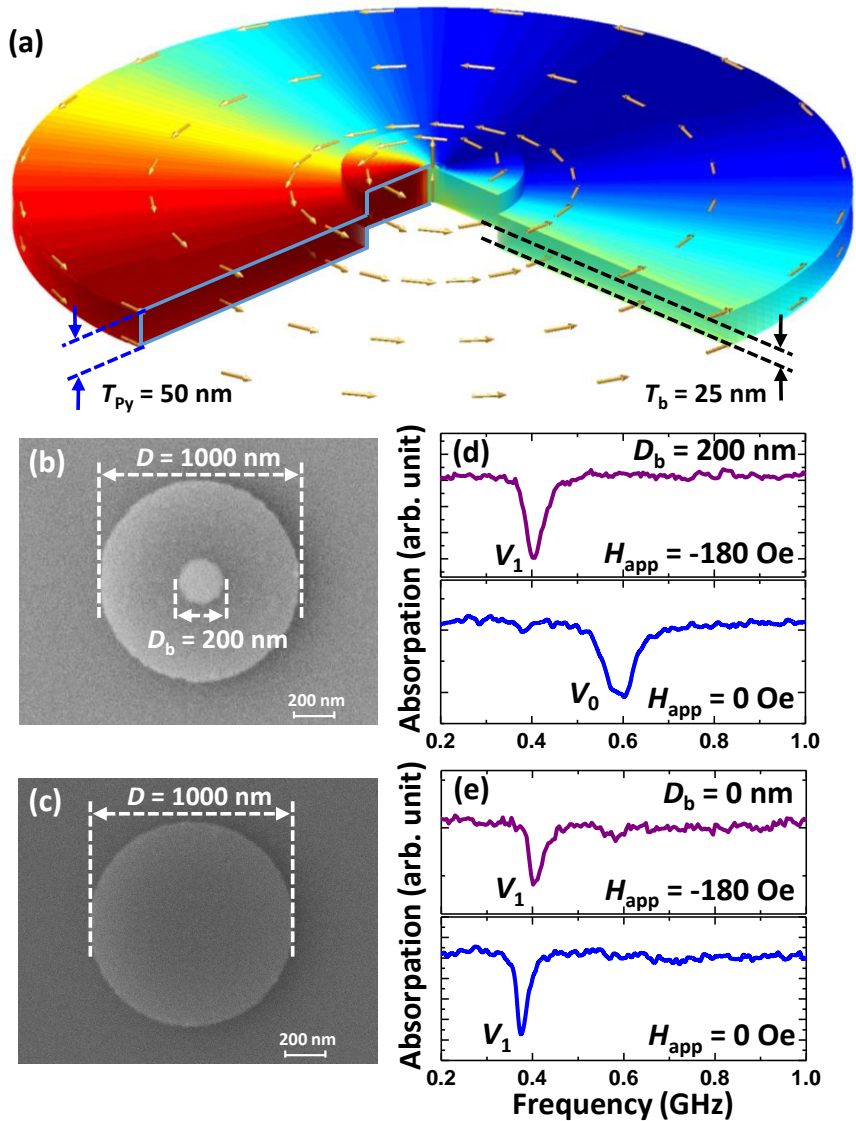

Fig. 1

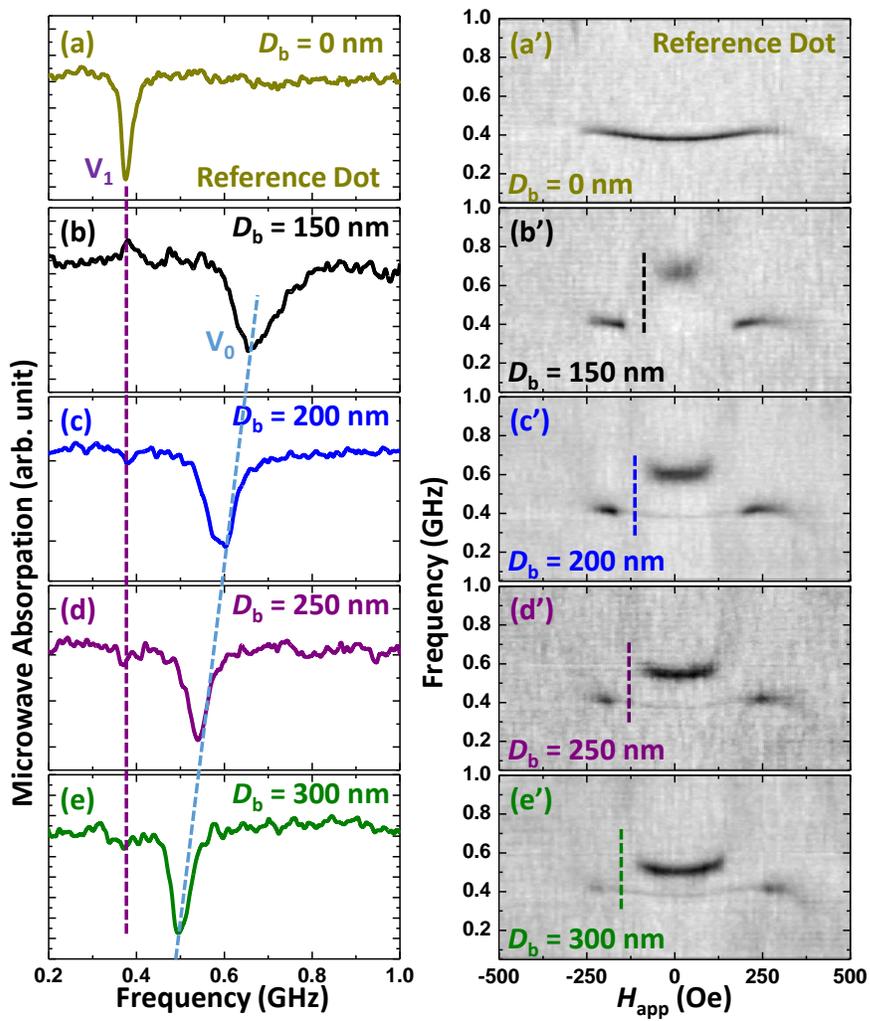

Fig. 2

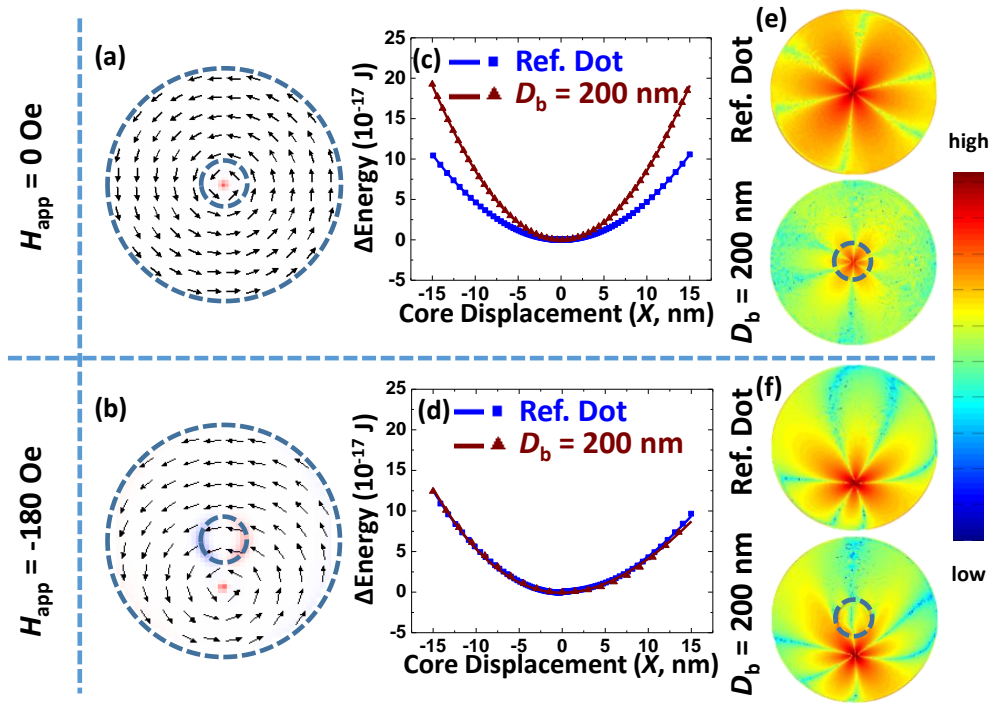

Fig. 3

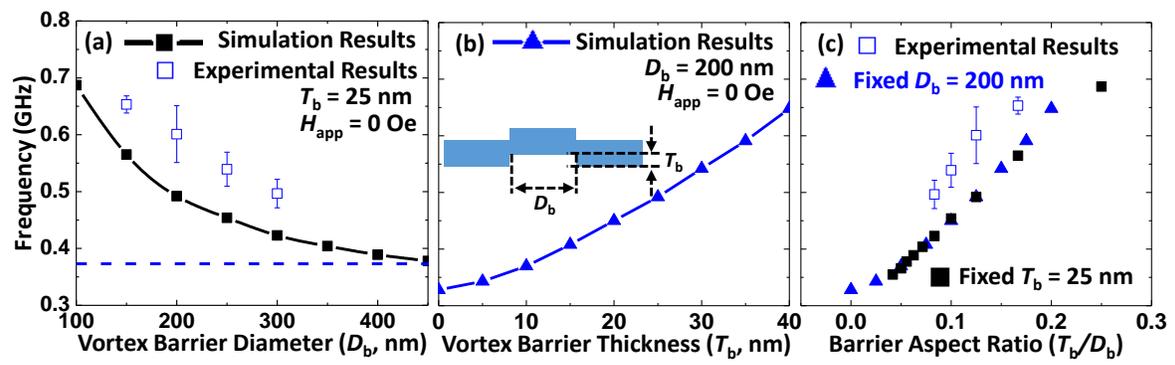

Fig. 4